\documentclass[usenatbib]{mn2e}
\usepackage{graphicx}
\usepackage{url}
\usepackage{lmodern}    
\usepackage[T1]{fontenc}

\newcommand {\andxxv}{And~XXII}
\newcommand {\andxxii}{And\,XXII}
\newcommand {\andxxiii}{And\,XXII}

\def\kms{km\,s^{-1}}
\def\kpc{kpc}
\def\FeH{Fe/H}

\def\ec4{EC4}

\def\ltsima{$\; \buildrel < \over \sim \;$}
\def\lta{\lower.5ex\hbox{\ltsima}}
\def\gtsima{$\; \buildrel > \over \sim \;$}
\def\simgt{\lower.5ex\hbox{\gtsima}}
\def\kms{{\rm\,km\,s^{-1}}}

\def\kpc{{\rm\,kpc}}

\def\msun{{\rm\,M_\odot}}
\def\lsun{{\rm\,L_\odot}}

\def\AA{$\; \buildrel \circ \over {\rm A}$}

\def\deg{^\circ}

\def\s{\ifmmode \widetilde \else \~\fi}
\def\={\overline}

\def\spose#1{\hbox to 0pt{#1\hss}}

\def\lta{\mathrel{\spose{\lower 3pt\hbox{$\mathchar"218$}}
     \raise 2.0pt\hbox{$\mathchar"13C$}}}
\def\gta{\mathrel{\spose{\lower 3pt\hbox{$\mathchar"218$}}
     \raise 2.0pt\hbox{$\mathchar"13E$}}}
\def\Dt{\spose{\raise 1.5ex\hbox{\hskip3pt$\mathchar"201$}}}    
\def\dt{\spose{\raise 1.0ex\hbox{\hskip2pt$\mathchar"201$}}}    

\def\dotsfill{\leaders\hbox to 1em{\hss.\hss}\hfill}

\def\FeH{{\rm[Fe/H]}}

\def\ec4{EC4}

\loadboldmathitalic

\title[]{Dynamics in the satellite system of Triangulum: 
\newline Is AndXXII a dwarf satellite of M33?}
\author[Scott\ Chapman et al.] 
{S.\ C.\ Chapman,$^1$ L.\ Widrow,$^2$ 
M.\ L.\ M.\ Collins,$^{3,1}$ 
J.\ Dubinski,$^4$ 
R.\ A.\ Ibata,$^5$ 
\newauthor
J.\ Pe\~narrubia,$^{6,1}$ 
M.\ Rich,$^7$
A.\ M.\ N.\ Ferguson,$^8$
 M.\ J.\ Irwin,$^1$   
G.\ F.\ Lewis,$^9$ N.\ Martin,$^{5,3}$ 
\newauthor
A.\ McConnachie,$^{10}$ 
N.\ Tanvir$^{11}$.
\\
$^{1}$ Institute of Astronomy, Madingley Road, Cambridge, CB3 0HA, U.K.\\
$^{2}$ Department of Physics, Engineering Physics and Astronomy, Queen's University, Kingston, Ontario B3H 3C3, Canada\\
$^3$ Max-Planck-Institut fur Astronomie, Konigstuhl 17, D-69117 Heidelberg, Germany\\ 
$^{4}$ Department of Astronomy \& Astrophysics, University of Toronto, 50 St George Street, Toronto, Ontario, Canada M5S 3H4\\
$^{5}$ Observatoire de Strasbourg, 11, rue de l'Universit\'e, F-67000, Strasbourg, France\\
$^{6}$ Ram\'on y Cajal Fellow, Instituto de Astrof\'isica de Andalucia-CSIC,
Glorieta de la Astronom\'ia, 18008, Granada, Spain\\
$^{7}$ Department of Physics and Astronomy, University of California, Los Angeles, CA 90095-1547\\
$^{8}$ Institute for Astronomy, University of Edinburgh, Royal Observatory, Blackford Hill, Edinburgh, UK EH9 3HJ\\ 
$^9$ Sydney Institute for Astronomy, School of Physics, A29, University of Sydney, NSW 2006, Australia\\ 
$^{10}$ NRC Herzberg Institute of Astrophysics, 5071 West Saanich Road, Victoria, British Columbia, Canada V9E 2E7\\ 
$^{11}$ Department of Physics \& Astronomy, University of Leicester, Leicester, LE17RH, UK\\
}

\date{\today}
\begin{document} 

\maketitle 
\begin{abstract} 
We present results from a spectroscopic survey of the dwarf spheroidal
\andxxii\ and the two extended clusters EC1 and EC2.  These three
objects are candidate satellites of the Triangulum galaxy, M33, which
itself is likely a satellite of M31.  We use the DEep Imaging
Multi-Object Spectrograph mounted on the Keck-II telescope to derive
radial velocities for candidate member stars of these objects and
thereby identify the stars that are most likely actual members.  Eleven
most probable stellar members (of 13 candidates) are found for \andxxii.  We obtain an upper
limit of $\sigma_v < 6.0 \,{\rm km\,s^{-1}}$ for the velocity
dispersion of \andxxii, [Fe/H]$\sim -1.6$ for its metallicity, and
$255\,{\rm pc}$ for the Plummer radius of its projected density
profile.  We construct a colour magnitude diagram for \andxxii\  and
identify both the red giant branch and the horizontal branch.  The
position of the latter is used to derive a heliocentric distance to
\andxxii\ of $853\pm 26 \,{\rm kpc}$.  The combination of the radial
velocity, distance, and angular position of \andxxii\  indicates that it
is a strong candidate for being the first known satellite of M33 and
one of the very few examples of a galactic satellite of a satellite.
N-body simulations imply that this conclusion is unchanged even if M31
and M33 had a strong encounter in the past few Gyr.  We test the
hypothesis that the extended clusters highlight tidally stripped
galaxies by searching for an excess cloud of halo-like stars in their
vicinity.  We find such a cloud for the case of EC1 but not EC2.  The
three objects imply a dynamical mass for M33 that is consistent with
previous estimates.

\end{abstract}
 
\begin{keywords} 
M\,33 -- M\,31 -- Dwarf Galaxies -- DEIMOS
\end{keywords}

\section{Introduction}

Dwarf spheroidal galaxies (dSphs) probe the outer regions of their host galaxies, and as such act as important tracers of the dark matter halo at large galacto-centric radii (e.g., Evans \& Wilkinson 2000, Watkins, Evans \& An 2010).
Further, dSphs are the progenitors of tidal streams found within the halos of the Milky Way (e.g., Belokurov et al.\ 2006, 2007) and Andromeda (e.g., Ibata et al.\ 2001, 2004, 2007; Chapman et al.\ 2006, 2008; McConnachie et al.\ 2009), which help to trace their orbits.  Modelling tidal
disruption of dwarfs provides another way to constrain the dark halos of their host galaxies.
DSphs can also be used as a test of our understanding of hierarchical clustering processes, including the 
evolution of these systems, and their survival in the environments of much larger dark matter haloes (e.g. Sales et al.\ 2007a,b).
Finally, the faint dSph population are believed to be dominated by dark matter, and can therefore provide useful laboratories to study the properties of dark matter itself (e.g., Pe\~narrubia et al.\ 2007,2008).


\andxxii, a faint dSph galaxy that lies in the Triangulum constellation, was discovered by Martin et al.\ (2009) using the Pan-Andromeda Archeological Survey (PAndAS -- McConnachie et al.\ 2009).
The 
proximity of \andxxii\  (42 kpc in projection) to the Triangulum galaxy (M33), in conjunction with its large distance from M31 (224 kpc, also in projection), led Martin et al.\ (2009) to speculate that 
this dwarf galaxy could in fact be the first discovered satellite dSph of M33.
However, at that point, no spectroscopic data were available for the dwarf, and neither the radial velocity nor a robust line of sight distance were known.

M33 may well have had a close passage by M31 and therefore \andxxii\ could
have been influenced by two more massive galaxies.
The existence of a satellite at this radius would place a constraint on the strength of an M33--M31 interaction (McConnachie et al., 2009, Dubinski  et al., in prep) 
since such an interaction naturally strips material from the outer parts of M33. 
If And\,XXII were shown to be a satellite of M33, then this would make M33 one of the
smallest galaxies known to have a satellite, and it would make
M31-M33-\andxxii\ one of the first examples of a three-level galaxy
hierarchy, a phenomenon that naturally arises in the hierarchical clustering scenario of structure formation, and a topic of considerable interest for predictions of cold dark matter (CDM) models
(Sales et al.\ 2007a; d'Onghia et al.\ 2010).
\andxxii\ would also expand the known halo size of the M33 system by 50\%, from its recently discovered outermost globular cluster (Huxor et al.\ 2009), and consistent with an extended stellar halo profile to large radius (Cockcroft et al.\ 2012). 


Assessing the number of expected bound satellites to a host galaxy of a given mass is not a straightforward exercise, as even moderate mass galaxies are heavily influenced by any larger galaxies they may have interacted with (e.g., Purcell et al.\ 2009).
M33 is a modest spiral galaxy with a mass (M$_{\rm dyn}\sim10^{11}$ M$_\odot$) approximately one tenth the mass of M31. It resides at a projected distance of $204\kpc$ from the centre of M31, presumably well-inside the M31 dark halo  and measurements of its line-of-sight velocity and proper motion suggest that it is on a highly elongated orbit around M31.  The implication is that M33 is a satellite of M31 and likely passed much closer to M31 in the past than its present distance (but see Loeb et al.\ 2005).  This association is supported by the existence of a strong warp and heavily disturbed  HI disk in M33 (Rogstad et al.\ 1976; Putman et al.\ 2009), 
as well as the presence of a stellar stream extending from its stellar disk (McConnachie et al.\ 2009, 2010). 
Interactions are likely important with galaxies like M\,33, 
influencing the number of bound dSphs. 

Through complete observations of a sample of nearby spiral galaxies, Erickson, Gottesman, \& Hunter (1999) typically find at least one massive satellite comparable to the LMC of spiral galaxies down to V$_{\rm r,max}$=140 km s$^{-1}$.
But the lowest mass galaxies in this survey are still more than twice as massive than M33, which has 
V$_{\rm r,max}\sim100$ km s$^{-1}$ from the stellar rotation curve out to 13~kpc (Trethewey\ 2011).
Guo et al.\ (2011) also attempt to characterize the satellite fractions in N-body simulations as a function of stellar mass, finding moderate mass galaxies like M33 found close to larger galaxies (like M31) 
are often deficient in bound satellites.
Extrapolating the Guo et al.\ (2011) results, M33 (with a stellar mass of 3-6$\times10^{9}$ M$_\odot$)  might be expected to have as many as 20 dSph satellites based on $\Lambda$CDM (Cold Dark Matter)  predictions, but in fact has only one candidate dSph out to a radius of 50~kpc projected, \andxxii, and this remains to be confirmed.
The lack of satellite dwarfs discovered around M33 
may suggest that it has been orbiting M31 for long enough to have had its satellites 
stripped by the more massive galaxy, and now indistinguishable from M31's satellites. 
 
While \andxxii\ is the only candidate dSph satellite of M33, there are other distant star clusters in the halo of M33 which may signpost, or even be the remnants of disrupted satellites.
The two `extended clusters', M33-EC1 
(Stonkute et al.\ 2008 -- hereafter EC1) and HM33-A (Huxor et al.\ 2009 -- hereafter EC2), lying in projection at 13~kpc
and 28~kpc respectively in M33's halo, represent just such a possibility.
These intriguing objects have no analogs in the Milky Way (MW) sub-system.
One extended cluster in  the M31 halo, EC4, has
been studied spectroscopically (Collins et al.\ 2009). EC4 lies within the `stream Cp'  tidal stream in M31's halo,
and the kinematics and metallicity of EC4 bear a striking resemblance to
`stream Cp' (Ibata et al.\ 2007; Chapman et al.\ 2008)
which has the same radial velocity and metallicity, 
suggesting that the two are  very likely related.
However, there
was no evidence that EC4 contained a substantial dark matter component to suggest it was the remnant of a dSph galaxy,
and EC4 was classified as a  stellar cluster with a large (30~pc)
core radius.  
The extended clusters (and outer halo clusters in general) have been found to
preferentially lie in regions of excess stars suggestive of tidal
debris in the halo of M31 (Mackey et al.\ 2010), and the large size
of ECs may be a byproduct of the disruption of their host galaxies.  In the
context of M33's dynamical mass, satellite system and accretion
history, it is thus of interest to study its EC1 and EC2, and
search for possible evidence of disrupted hosts in their vicinity.

Whether \andxxii, as well as EC1 and EC2, are likely to orbit M33  depends on their kinematics and real distance 
from the two possible hosts. 
These issues can be addressed  via the radial velocities of the brightest 
Red Giant Branch (RGB) stars, which in turn confirm whether the  brightest 
stars  are likely to be members. 
Tollerud et al.\ (2012) presented  radial velocities of 7 stars in \andxxii\ as part of their survey of M31 dSphs, noting that it might be a candidate for a satellite of M33.
The distance to \andxxii\ can also be corroborated and more precisely constrained by  photometry reaching the stellar horizontal branch. 
In conjunction with simulations investigating the bound particles and  
the gravitational forces of M31 and M33 on \andxxii, we can attempt to determine whether \andxxii\ is an M33 satellite. 

To address the question of \andxxii's parent galaxy and the detailed
properties of the dSph itself, we have analyzed the optical imaging from
PAndAS (McConnachie et al.\ 2009), and used the DEep Imaging
Multi-Object Spectrograph (DEIMOS) on Keck~II to derive radial
velocities and metallicities of stars within \andxxii.  
To complete
the picture of satellites of M33, we have also obtained spectroscopic
observations of  EC1 and EC2 (two likely additions to the M33 satellite system),  in order to establish their radial velocities and constrain the halo mass of M33, as well as search for the signatures of dark matter or tidal streams associated with them.
In this work, we assume heliocentric distances to M31 and M33 respectively of
779$\pm$19~kpc and 820$\pm19$~kpc, (Conn et al.\ 2012), consistent
with distances to M33 previously estimated (e.g., McConnachie et
al.\ 2004, Tiede et al., 2004).

\begin{figure}
\centering
\includegraphics[width=0.5\textwidth, angle=0]{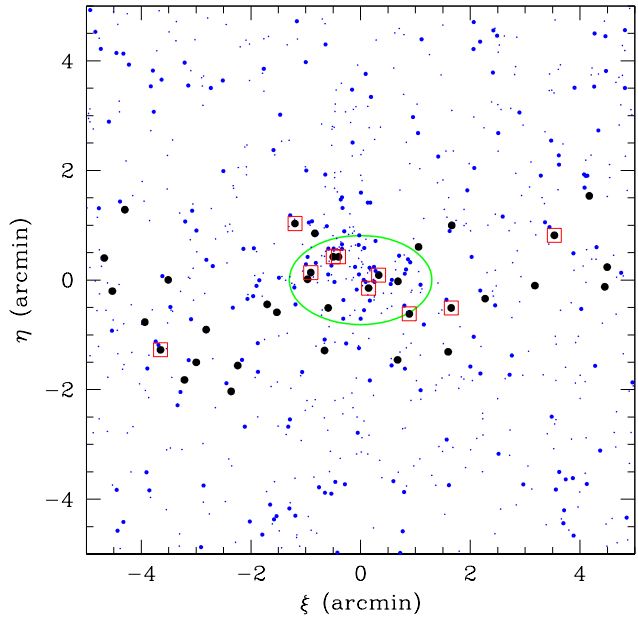}
\caption{Spatial distribution of stars in the region of  \andxxii.
All stars in the 
PAndAS survey (small blue dots) contrast the centrally concentrated subset  with colours and 
magnitudes consistent with metal-poor red giant branch populations at the distance of M31 (larger blue dots). All stars lying in the DEIMOS mask are shown (large black filled circles),
while candidate member stars  identified spectroscopically are highlighted (red squares).
Member stars were identified probabilistically 
as described in text and Collins et al.\ (2012).
The central  ellipse corresponds to the region within two half-light radii of the dwarf galaxy 
using structural parameters listed in Table 1. 
\label{a22xieta}}
\end{figure}

\begin{figure}
\centering
\includegraphics[width=0.53\textwidth, angle=0]{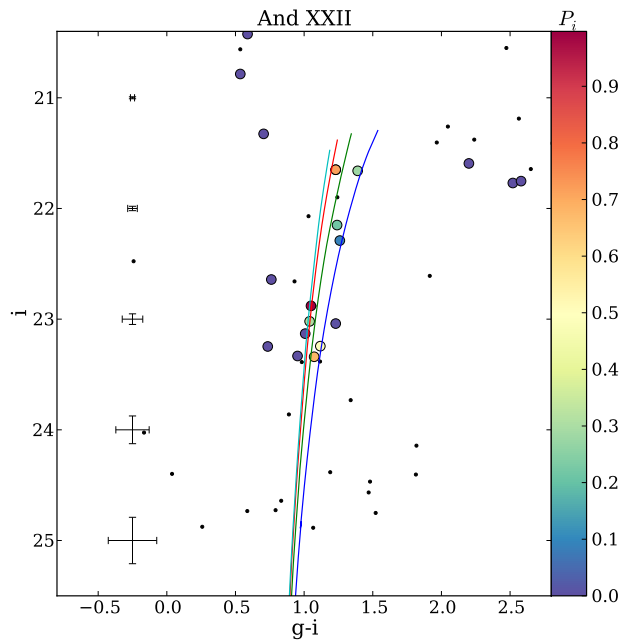} 
\caption{CFHT-MegaCam colour-magnitude diagrams of stars 
showing the DEIMOS targeted stars (large symbols) and all star-like sources within 2$'$ radius (small symbols) centered around \andxxii. Dartmouth isochrones, from [Fe/H]= $-1.8$ to $-1.4$ are overlaid at our derived distance to \andxxii.
 The RGB is clearly visible, while the horizontal branch is just visible  although without any likely spectroscopic members (the horizontal branch is however well detected in the $g-$band -- Fig.~\ref{HB}). 
 Candidate dSph member stars from the DEIMOS spectroscopy are further
highlighted (large circles, colour-coded by their probability).  
Probability of membership to \andxxii\  is defined by a combination of velocity, distance from the RGB locus, and radial distance from the centre of \andxxii\ (see Table~2).
The NaI doublet equivalent width was used as an additional discriminant of MW foreground stars, however this did not
result in the rejection of any stars not already flagged from the CMD position.
\label{a22cmd}}
\end{figure}

\begin{figure}
\centering
\includegraphics[width=0.5\textwidth, angle=0]{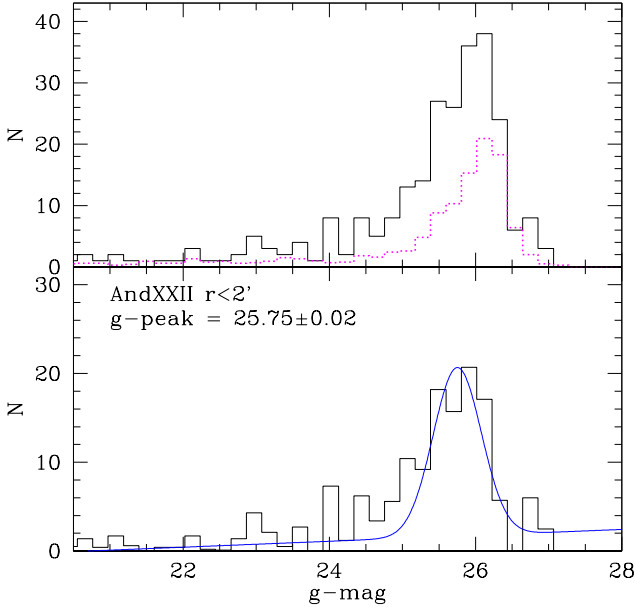}
\caption{Top panel: The $g$-band luminosity function of the 
central regions of \andxxii\ (black) within two half-light radii, together with suitably scaled (by sky area)
much larger neighbouring comparison regions (magenta). 
No $g-i$ cut has been applied given the faintness of the HB stars.
Bottom panel: Difference luminosity function highlighting the regions around the horizontal 
branch. 
The $g$-band magnitude of the peak of the fits is consistent at $g=25.75\pm0.10$
corresponding to a distance of $853\pm26$~kpc, with a measured E(B-V) of 0.07 extinction in this region.
The result is insensitive to a range of radii (from r$_{1/2}$=1--3 arcmin) used for the luminosity function.
\label{HB}}
\end{figure}

\begin{figure}
\centering
\includegraphics[width=0.5\textwidth, angle=0]{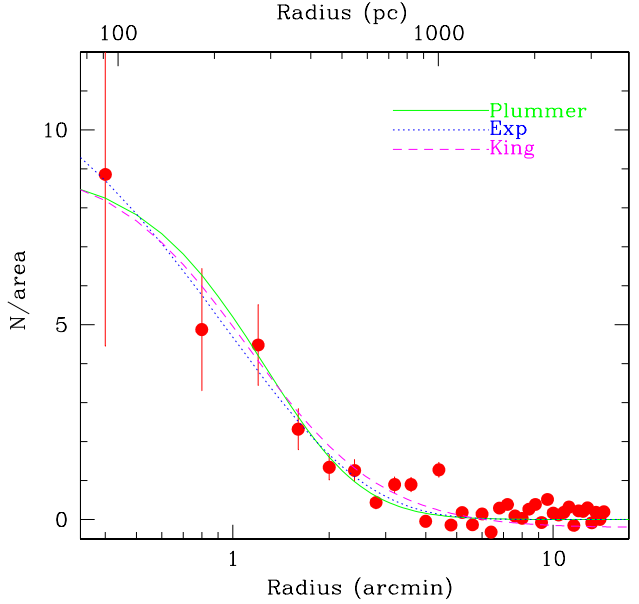}
\caption{Radial profile fits from CFHT-MegaCam extracted stars  (with point-like stellar profiles) using 0.4'  elliptical annular ring widths for binning, with background subtracted. 
Using stars with $20.5<I<24.0$ stars we derive r$_h$(Plummer)=1.0$\pm$0.3 arcmin.
\label{a22profile}}
\end{figure}

\begin{figure}
\centering
\includegraphics[width=0.5\textwidth, angle=0]{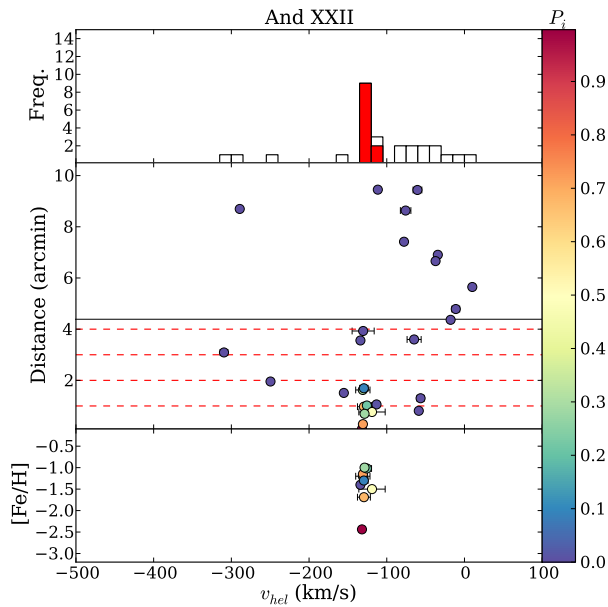} 
\caption{
Distribution of stars in \andxxiii\ as a function of their radial velocity
(upper panel). 
The systemic velocity of M33 (-178 $\kms$) is much closer to \andxxii\ than that of M31 (-300 $\kms$).
The stars are then shown as a function of radius 
from the centre of \andxxii, with multiples of the half-light radius drawn as a dashed lines (middle panel). Stars are colour-coded by their membership probability, as in Fig.~2.
Photometric [Fe/H] as described in the text is shown on the bottom panels, 
revealing the tight range in metallicities of \andxxiii\ ([Fe/H]$\sim$-1.6). Most of the foreground Milky Way stars lie off the grid of isochrones used to calculate [Fe/H].
\label{a22vels}}
\end{figure}

\begin{figure}
\centering
\includegraphics[width=0.53\textwidth, angle=0]{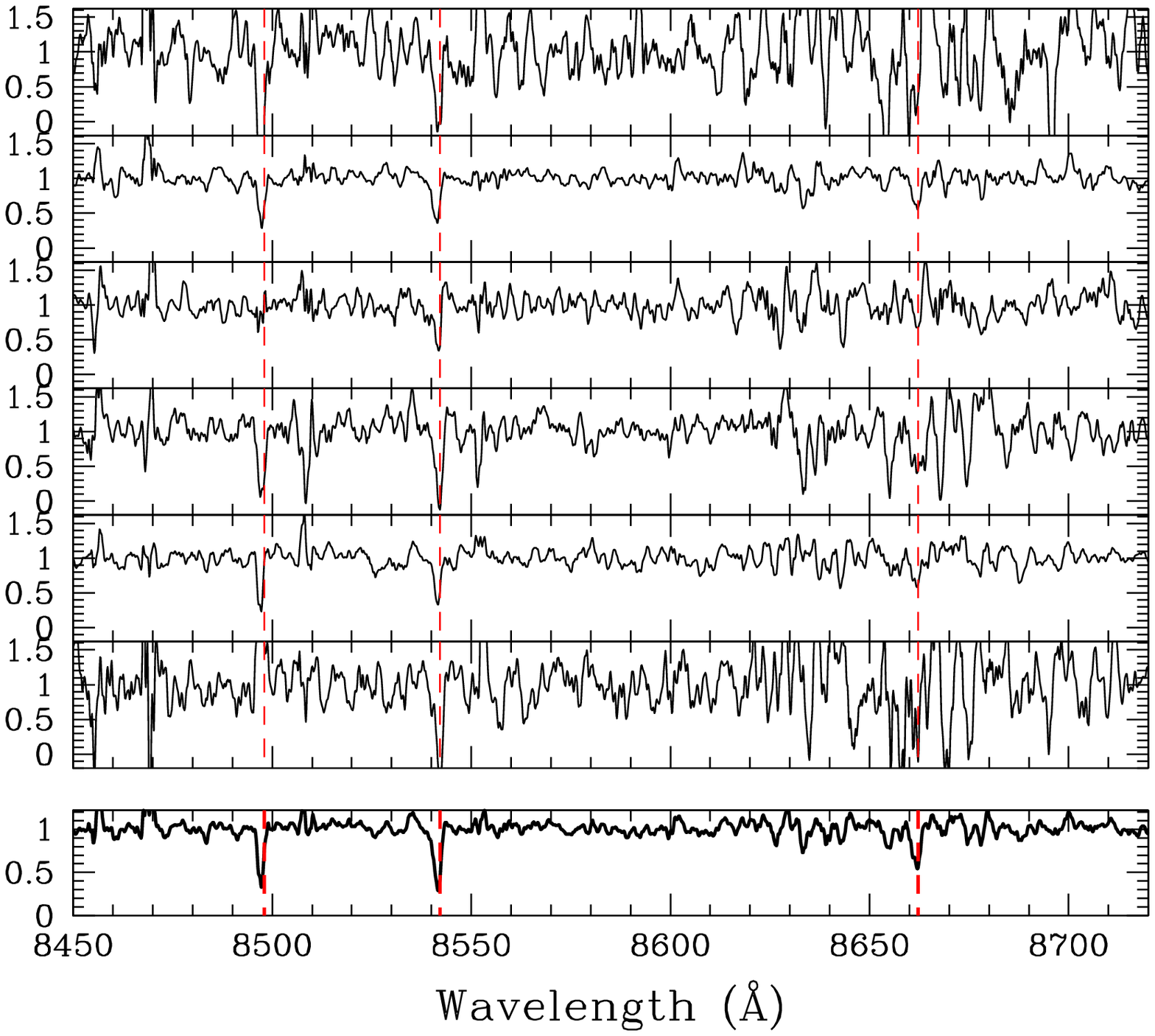}
\vskip-1.1cm
\caption{Spectra of the six highest probability \andxxii\ member stars, with probability of membership $>$0.3, see Collins et al.\ (2012) for details.
The spectra emphasize the well detected CaII triplet lines in each case.
The inverse 
variance weighted summed spectrum is shown in the bottom offset 
panel. 
\label{a22spec}}
\end{figure}

\section{Observations}
Multi-object Keck observations with DEIMOS (Faber et al.\ 2001) 
for \andxxiii\
were made on 2009 Sept.\ 23 and 2010 Sept.\ 9.
The first set of observations had 0.8$''$ seeing and total mask exposure times of 20min while the second set had 0.6$''$ seeing and 60min exposure times.
The extended clusters, EC1 and EC2, were also observed on 2009
Sept.\ 23 for 45min each.  We use the 1200 line/mm
grating where the spectral resolution in the red is R$\sim$6000
and the observed wavelength range is $0.65-0.93 \mu$m.
Data reduction for the spectra followed standard techniques using the
DEIMOS-DEEP2 pipeline (Faber et al. 2003) and included debiassing,
flat-fielding, extracting, wavelength-calibration and sky-subtraction.
The data were also reduced using our custom pipeline (e.g., Ibata et
al.\ 2005) as a check on the extractions and calibrations.

Radial velocities of our target stars were determined by fitting the
peak of the cross-correlation function between the observed spectra
and a template spectrum.  The latter consisted of delta functions
smoothed to the instrumental resolution (1.3\AA) at the wavelengths of
the Ca{\sc II} triplet (CaT) absorption lines. This procedure also
provided an estimate of the radial velocity accuracy obtained for each
measurement. The velocity uncertainties range from $4$ to $17 \kms$.

Target stars were assigned from a broad (1 mag) box around the general
outline of the RGB of the dwarf.  A higher priority was given to stars
within a narrower box of 0.3 mag around the RGB.  In addition,
brighter stars were prioritized over fainter ones.
The remainder of the mask space was filled with target stars from a
broad region that encompassed RGB stars in M31 over all possible
metallicities and distances.  The \andxxii\ masks had 93 and 173
target stars respectively, while EC1 had 168, and EC2 had 88 target
stars.

We followed the procedures outlined in Collins et al.\ (2010)
in order to correct for various systematic errors that arise in the
velocity calculations.
A template telluric spectrum was constructed from the data, which was
cross-correlated with each of our science spectra to determine the
velocity shift caused by systematic errors. For brighter stars
($i<21$) 
the velocity corrections are approximately constant.  However for
fainter stars, where the Signal-to-Noise (S:N) is lower, the magnitude
of this telluric correction shows significant scatter about the median
value.  Evidently, this effect is injecting noise into the velocity
measurements.  We apply a correction corresponding to a 2.1$\kms$
offset and a small (1.9$\kms$ mask end to end) velocity gradient.
This correction is based on measurements of the $i<21$ stars in the
mask and presumably accounts for a slight mask rotation and positional
offset of stars in the slits.

\begin{figure}
\centering
\includegraphics[width=0.5\textwidth, angle=0]{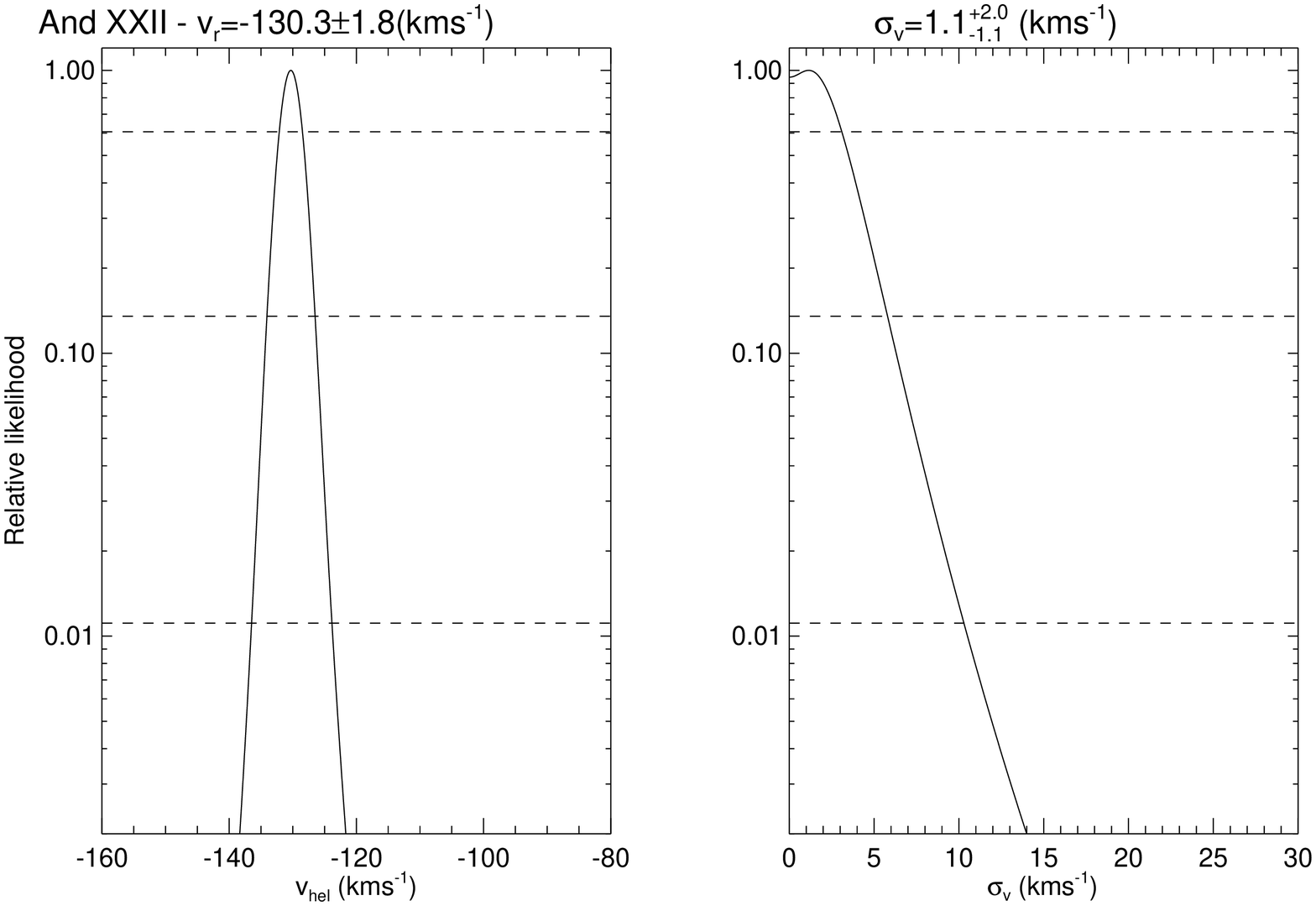}
\caption{Likelihood distributions of member stars in \andxxv, showing the dispersion is essentially 
unresolved 
with v$_r = -130.3\pm1.8 \kms$, 
and $\sigma_v < 4.0 \kms$, 95\% confidence.
\label{a22ML}}
\end{figure}

There are twenty-four stars for which the velocity was determined
with both of our masks.  These repeat observations provide us with 
a further handle on the velocity measurement errors.  In principle, the quantity
\begin{equation}
D_i\equiv \frac{v_{1,i} - v_{2,i}}{\left (\sigma_{1,i}^2 + \sigma_{2,i}^2\right )^{1/2}}
\end{equation}
should be a Gaussian random variable with unit variance.  
Here, the subscript $1,i$ refers to the first measurement of the $i$'th star,
etc.  Likewise, $\sigma_{1,i}$ is the Monte Carlo error estimate for
that measurement.  Our procedure is to replace the denominator
by $\left (\sigma_{1,i}^2 + \sigma_{2,i}^2 + \epsilon^2\right )^{1/2}$
and adjust the free parameter until the desired distribution is achieved.
We find $\epsilon=1.6\,{\rm km\,s}^{-1}$, comparable to that found in
  other work (e.g. $2.2\,{\rm km\,s}^{-1}$ in {Simon \& Geha\ 2007}).
    We conclude we can measure velocities to accuracies of $\sim
    5\,{\rm km\,s}^{-1}$ down to a S:N of 3 or better.

\section{Results -- And\,XXII}

\begin{table}
\caption{Derived properties of \andxxii}
\label{parameters}
\begin{tabular}{l|cc}
\hline\hline
$\alpha$ (J2000) & $01^{\rm h}27^{\rm m}39.9^{\rm s} \pm 0.6^{\rm s}$ \\
$\delta$ (J2000) & $28\deg05'28'' \pm6''$ &  \\
$D$ (kpc) & $853\pm26$ \\
$r_\mathrm{M31}$ (kpc) & $258\pm31$  \\
$r_\mathrm{M33}$ (kpc) & $59^{+21}_{-14}$ \\
$v_{r} (\kms)$ & $-130.0\pm1.7$\\
$\sigma_{v} (\kms)$ & $<6.0$, 99.5\% confidence\\
$\langle\FeH\rangle_{phot}$ & $-1.58\pm0.04$  \\
$\langle\FeH\rangle_{spec}$ & $-1.62\pm0.05$  \\
$r_h$ (arcmin) & $1.0\pm0.3$  \\
$r_h$ (pc) & $255\pm78$ \\
PA (NtoE) ($\deg$) & $93\pm13$ \\
$\epsilon=1-b/a$ & $0.55\pm0.10$ \\
$L_V$ ($\lsun$) & $0.35\pm0.05\times10^5$  \\ 
$\mu_{V,0}$ (mag/arsec$^2$) & $26.7\pm0.6$ \\
\hline
\end{tabular}\\
\end{table}

The spatial distributions of the stars in the region of \andxxii\ is
shown in Figure~\ref{a22xieta}. A central concentration of stars that have
colours and magnitudes consistent with a metal-poor RGB population at the distance of M33 is just visible by eye.
Spectroscopically targeted stars and candidate member stars (discussed below) are
highlighted in  Figure~\ref{a22xieta} and listed in Table~2.  %
In this section, we first analyze the photometry for \andxxii, to assess a distance and measure structural parameters. We then use our spectroscopic survey to constrain candidate member stars and derive the kinematic properties.

\subsection{Photometry}

\subsubsection{Distance}

The colour-magnitude diagram for \andxxii\ is shown in Figure~\ref{a22cmd}.
The \andxxii\ dwarf galaxy is too poorly populated to directly
determine its distance using traditional tip of the red giant branch
(TRGB) methods, in spite of the fact that our kinematic measurements
have secured the brightest RGB stars as dwarf members.  Martin et
al.\ (2009) effectively avoided calculating the TRGB distance
entirely, while Conn et al.\ (2012) reveal a ML-TRGB distance with
large errors (922$^{+36}_{-148}$~kpc).  To complement the
crude TRGB distance, we use the reprocessed CFHT Megacam photometry
and exploit the extra depth of the $g$-band for bluer stellar
populations to directly measure the luminosity of the horizontal
branch (HB). This feature accounts for the strong over-densities
sloping to fainter magnitudes and bluer colours at the bottom of the
Colour-Magnitude Diagram (CMD), e.g. from $g-i =1.0, i=24.8$ to $g-i =0.0$, $i =25.3$ 
The $i$-band depth cuts through this region precluding direct isochrone 
fitting, but we can still exploit the better depth of the $g$-band 
directly by analyzing the $g$-band luminosity function instead. As a 
 benchmark, the HB magnitude of M31 is $g_0=25.2$, which for the average
extinction in the halo of M31 translates to an apparent magnitude of
$g=25.4-25.5$. The $g$-band data extends down to
$ g < 27.0$ mag, where the comparison regions in the CMDs show that there is
likely a notable component of contamination from unresolved background
 galaxies, and careful subtraction of the comparison regions is required
to measure the HB mag.
The observed $g$-band luminosity function (LF) is shown in
Fig.~\ref{HB} along with those of the appropriately scaled (by area)
background reference field (from a square degree surrounding \andxxii).  The bottom panel of Fig.~\ref{HB} show
the $g$-band LF with the background subtracted and a best fit Gaussian
magnitude.  To calculate the distance modulus we must first estimate a
standard absolute magnitude, M$_g$ for the HB populations typical of
dSphs. We adopt a constant of $M_g = 0.8\pm0.1$ in the CFHT MegaCam AB
magnitude system, as derived in Richardson et al.\ (2011). Empirical
models which estimate the closely related M$_v$ of the HB
(e.g.\ Gratton 1998) have been predominantly determined from globular
clusters, and have a weak metallicity dependence (gradient) ($<$0.1
mag at [Fe/H]$\sim$-1.6).  The $g$-band magnitude of the peak of the
fit is $g=25.75\pm0.10$, and the extinction corrected distance is
853$\pm$26~kpc, with the error  calculated by propagating the error in M$_g$ , the error in the measured location of the HB, and the $\pm$3\% uncertainty in the calibration of the photometry. 
The luminosities and sizes are updated from Martin et
al.\ (2009) accordingly in Table~1.

\subsubsection{Structural parameters}

Martin et al.\ (2009) measured a radial profile for \andxxii\ using
CFHT Megacam data.  Here, we take advantage of improvements in the
CFHT image processing to reassess the structural properties of
\andxxii.  We first re-estimate the ellipticity using our magnitude limit $i<24$.
Background-corrected radial profiles of \andxxii\ (Figure~\ref{a22profile})
are then measured using the average stellar density within series of fixed
elliptical annuli using the parameters from Table~1.
The error bars account for Poisson counting statistics and uncertainties 
in the derived background level. The overlaid model curves are 
fits to the data.  The best-fit Plummer scale length is
r$_h$(plummer)=1.0$'$$\pm$0.2$'$ when we use stars with $20.5<i<24.0$.
We find that varying the catalog depth from $i_{\rm max}=23$ to
$i_{\rm max}=24.5$, or using the deeper $g$-band data has little
effect on the fit.  The two half-light radius limit is shown in
Figure~\ref{a22xieta}.  This radius is slightly larger than that found in Martin et
al.\ (2009) (0.94$'$$\pm$0.2$'$).  With our new distance estimate
(853~kpc), we constrain the physical size of
\andxxii\ to be 255$\pm$78~pc.

\subsection{Spectroscopy}

\subsubsection{Membership}
A  CMD of stars derived from the CFHT-MegaCam imaging within a two
arcminute radius of \andxxiii\ is shown in Figure~\ref{a22cmd}.
Spectroscopically targeted stars and likely member stars
are highlighted. 
The red giant branch is clearly visible, while in the $g$-band data,  even the horizontal
branch is well detected (Figure~\ref{HB}).
The velocity distribution of stars in And\,XXII is shown in Figure~\ref{a22vels},
highlighting the clear peak of 10 stars near $-130 \kms$, representing
\andxxii\ (spectra are shown in Figure~\ref{a22spec}). 
The stars are also shown as a function of
radius from the centre of the dwarf, with multiples of the half-light
radius indicated.  The probability that a given star is a member of
\andxxii\ is determined from a combination of its radial velocity, its
distance from the RGB locus in the CMD, and its radial distance from
the centre of \andxxii. The motivation and methodology of the membership probability assignment 
 is described in detail in Collins et al.\ (2012).  These probabilities
are listed in Table~2 for member stars, and reveal that the two
outlying candidate stars at four half-light radii have a very low chance
of being members of the dSph, but the others are likely to be members.

\subsubsection{Radial velocity and dispersion}

The maximum likelihood approach provides a method to assess the true
underlying velocity distribution of the dSphs even when the velocity
errors for member stars are large and variable (see, for example,
Martin et al.\ 2007), which is the case here. In our analysis, we
weight each member star by its probability (Table~2), as described in
detail in Collins et al.\ (2012). Figure~\ref{a22ML} shows the likelihood
distribution as a function of the mean motion $v_r$ and velocity
dispersion $\sigma_v$ of \andxxii.  The dispersion is not quite
resolved at the $1\sigma$ level.  We find v$_r$ =-130.0$\pm1.7 \kms$
and $\sigma_v<6.0 \kms$\ at 99.5\% confidence,  formally
consistent with zero.  This in line with other M31 dSphs (see,
for example, Tollerud et al.\ 2012, Collins et al.\ 2010, 2012).
Specifically, Tollerud et al.\ (2012) find v$_r$ =-126.8$\pm3.1\kms$
and $\sigma_v=3.54^{+4.16}_{-2.49} \kms$ from a sample of seven stars, in agreement with our findings.

Photometric metallicity, [Fe/H]$_{\rm phot}$, is estimated by comparison
to Dartmouth isochrones (Dotter et al.\ 2008) corrected for extinction
and shifted to the distance of \andxxii.  This [Fe/H] value is shown
in Figure~\ref{a22vels} as a function of radial velocity, revealing the tight range
in metallicities for most stars in \andxxii\ (median [Fe/H]=-1.6,
interquartile range $\pm$0.1).

One-dimensional spectra of the most robust member stars in
\andxxiii\ are shown in Figure~\ref{a22spec}. All spectra show CaT 
cross-correlation peak $> 0.2$, and lie on the well defined RGB of
\andxxiii. The inverse variance weighted, summed spectrum is shown in
the bottom offset panel.  While not shown in the spectra, the Na{\sc
  I} doublet is significantly undetected  in the individual stars, and
also in the stacked spectrum. 
The Na{\sc I} equivalent width is
sensitive to the surface gravity, and thus is a good discriminant of
Galactic foreground dwarf stars (Schiavon et al.\ 1997). 
Stars with velocities $> -100
\kms$, typical  Galactic star velocities, often have well detected
Na{\sc I} doublet lines in our DEIMOS spectra. 
At the velocity of \andxxii, $\sim -130 \kms$, the typical fraction of Galactic stars has already dropped close to zero
(e.g., Collins et al.\ 2012),
and only one star in our CMD selection box, lying within two core
radii of And\,XXII, is obviously Galactic with strong Na{\sc I} equivalent width.  
The spectroscopic [Fe/H] has very large errors in most of the individual spectra, however a reasonable comparison can be made between the [Fe/H] derived from the  stacked spectrum and the median photometric [Fe/H]. We find  [Fe/H]$=-1.62$ by measuring the EW of the Ca{\sc II} triplet lines as in Chapman et al.\ (2005), on the \citet{carretta} scale. 
Koch et al.\ (2008) have also demonstrated that high quality Keck/DEIMOS spectra of stars in the M31 halo are amenable to further chemical analysis, showing a range of species (mostly Fe{\sc I} and Ti{\sc I} lines) which become weaker for the more metal-poor stars.
Our stacked spectra do not detect significant absorption at the Ti{\sc I} lines
(8378, 8426, and 8435\AA), but consistent with the expected equivalent widths for stars of similar metallicity from Koch et al.\ (2008). 

\section{Results -- EC1 and EC2}

We next turn our attention to the two extended clusters EC1 and EC2,
which were recently discovered in the halo of M33.  As discussed in the
introduction, extended clusters (Huxor et al.\ 2005, 2008, 2009) could have several viable explanations:
compact, tidally perturbed dwarf galaxies,  clusters associated to stellar debris from low
surface brightness dSphs that have been disrupted, or simply large clusters.

\begin{figure}
\centering
\includegraphics[width=0.45\textwidth, angle=0]{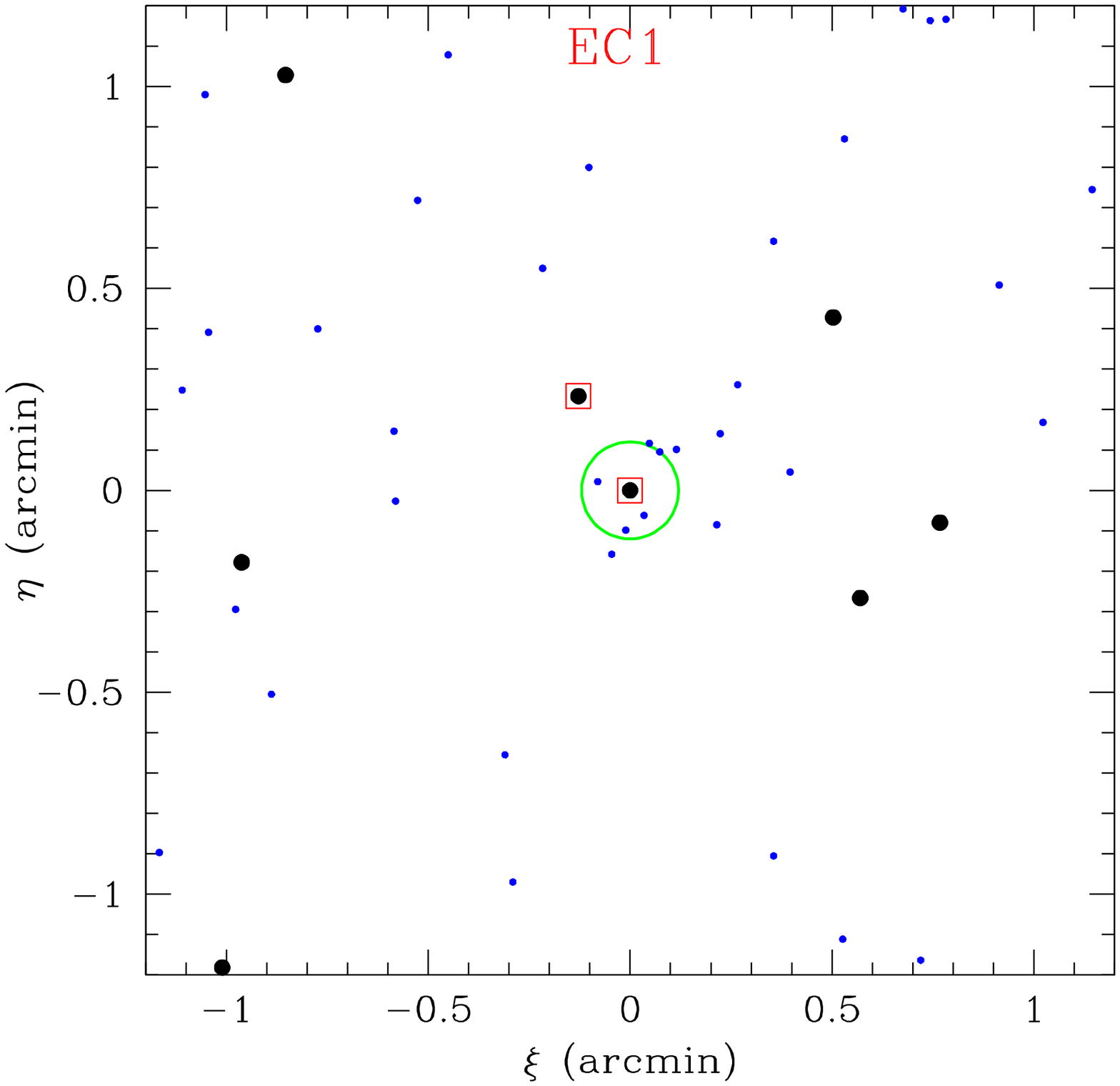}
\includegraphics[width=0.45\textwidth, angle=0]{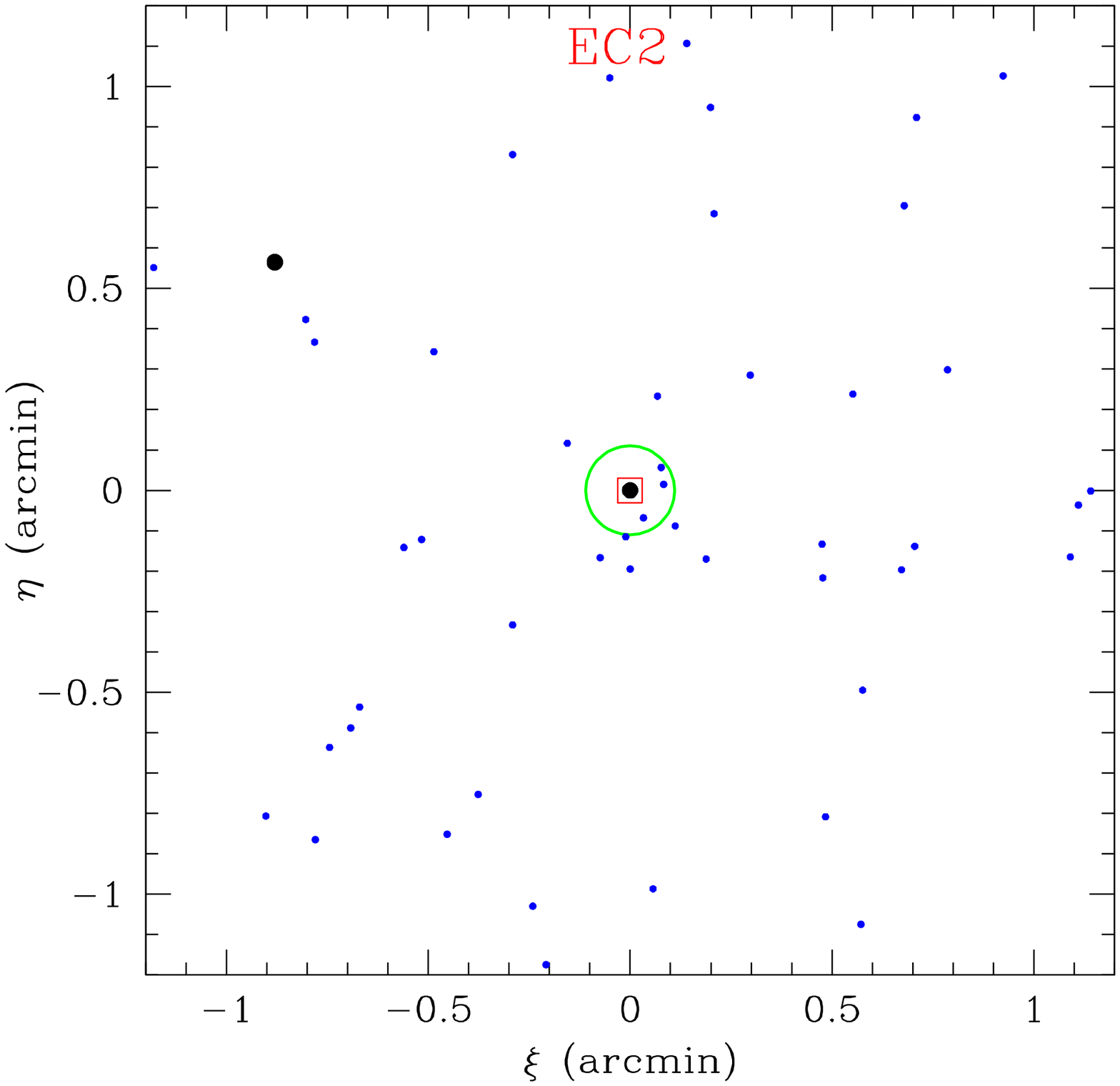}
\caption{ Spatial distribution of stars in the regions arounds EC1
  (top) and EC2 (bottom).  Stars with colours and magnitudes consistent
  with metal-poor red giant branch populations at the distance of M33
  (blue dots) reveal the small concentrations associated to EC1 and
  EC2. Blending in the central cluster regions results in not all
  cluster stars being extracted by our cataloging (as depicted above). All stars lying in
  the DEIMOS mask are shown (large black circles), while candidate
  member stars identified spectroscopically are highlighted (red
  squares).  The central circle highlights the core radii in each case.
\label{ecxieta}}
\end{figure}

The M33 extended clusters, EC1 (Stonkute et al.\ 2008) and EC2 (Huxor et al.\ 2009), are neither as
luminous, nor as extended as the extended cluster EC4 
in the halo of M31, studied in detail by
Collins et al.\ (2009).  We obtained spectra
for only a few RGB stars in M33-EC1 and EC2 since most of the stars in
these systems  were either too faint or  too crowded in the field to permit multiple slits  being placed.
In Figure~\ref{ecxieta}, we show the stars in the $5$ arcmin$^2$
field around each cluster,  although blending in the central cluster regions results in not all
cluster stars being extracted by our cataloging -- the reader is referred to the EC1 HST data in Stonkute et al.\ (2008) . 
The figure shows the small core radii for the two ECs (20.3 and
18.0~pc respectively) and the dearth of stars suitable for
spectroscopic followup.  While a spectroscopic slit is placed over the
brightest star in each EC, along with fainter neighbouring stars on
the slit, the next spectroscopically accessible bright star is beyond the core radius in
each case.  With a single spectroscopic mask, it is therefore
difficult to study these faint, relatively compact systems.

\begin{figure}
\centering
\includegraphics[width=0.45\textwidth, angle=0]{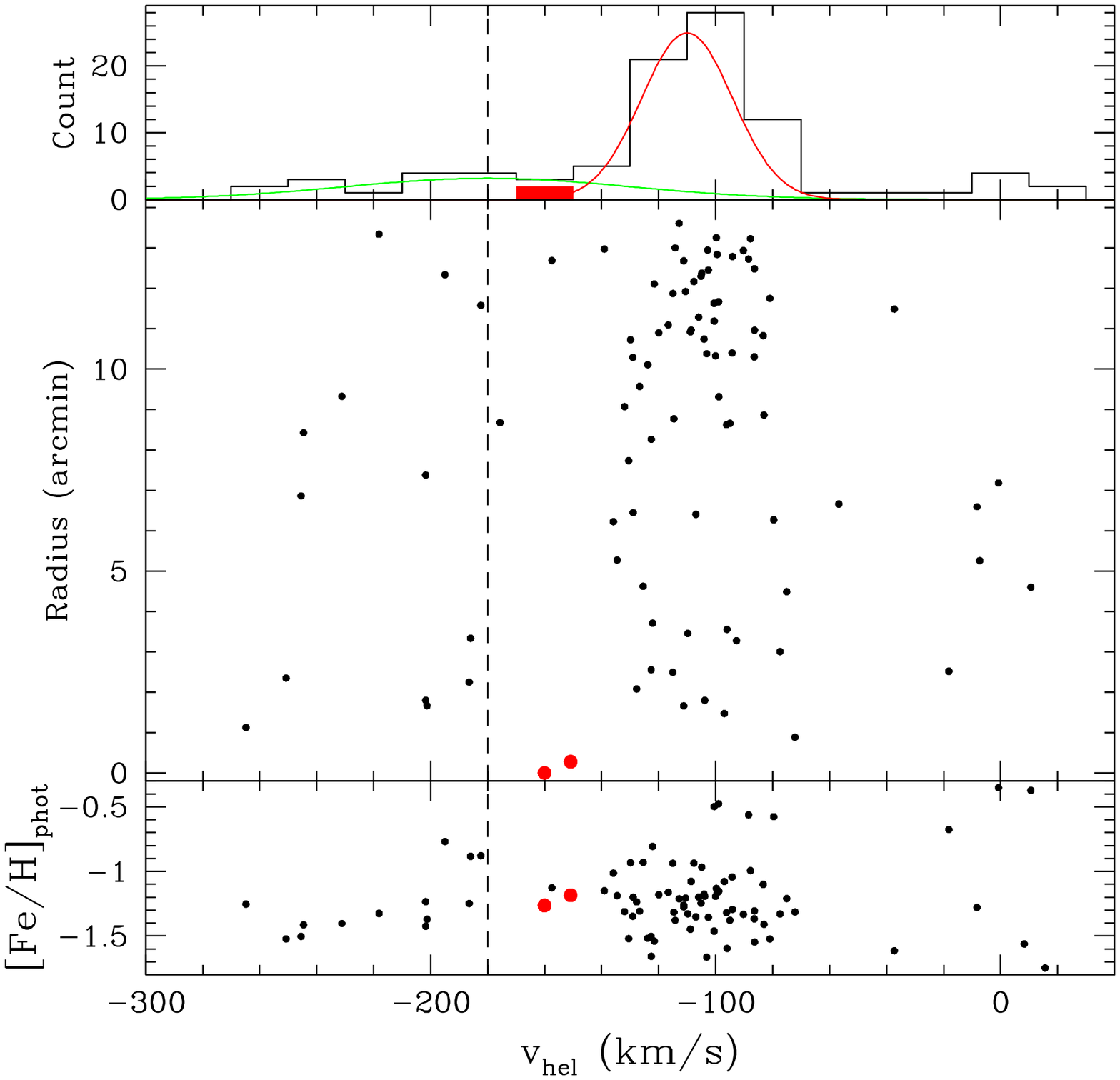}
\includegraphics[width=0.45\textwidth, angle=0]{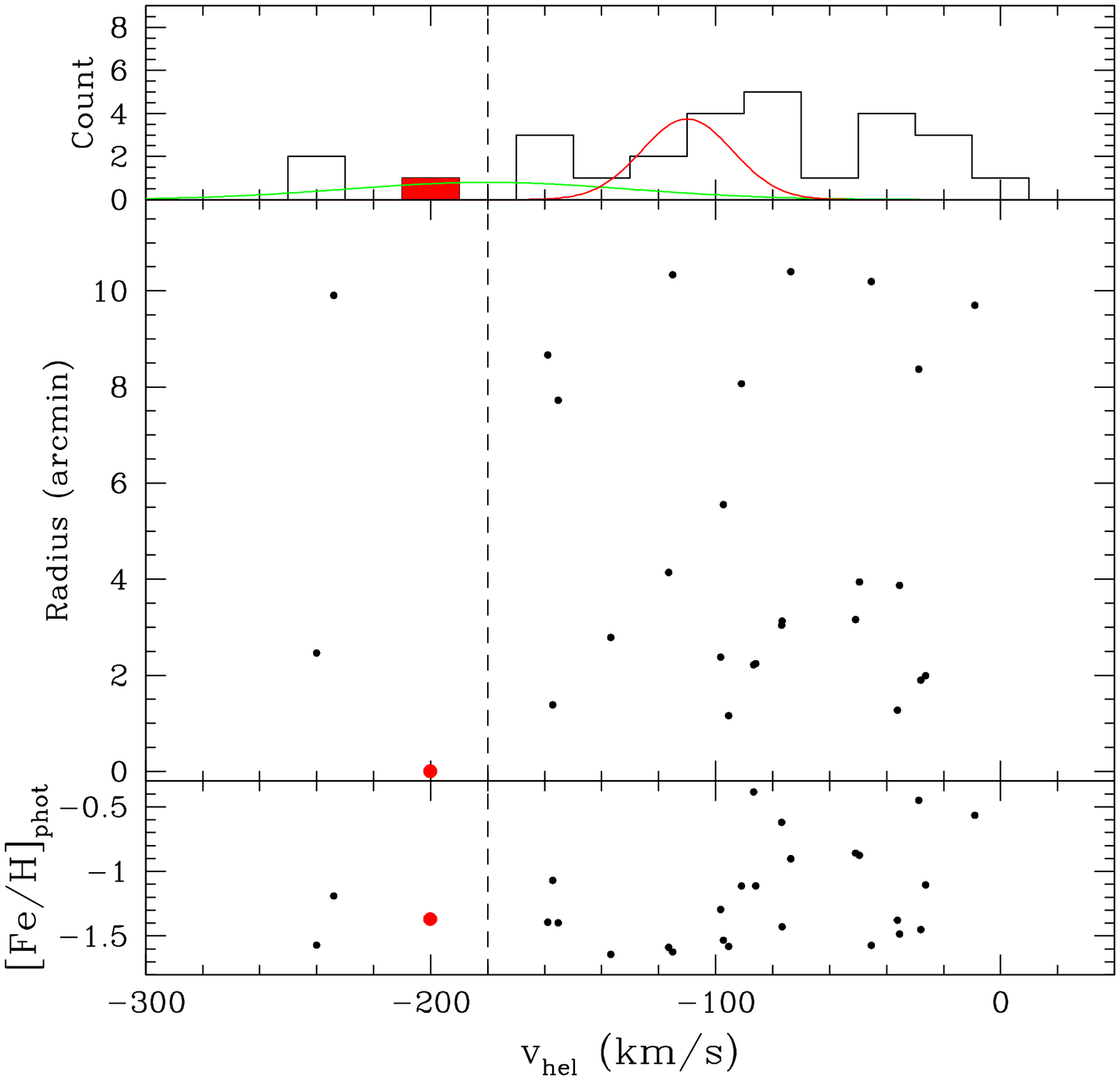}
\caption{
The distribution of stars in the fields of EC1 (top) and EC2 (bottom)  as a function of their radial velocity, where two likely member stars are found for EC1, and only one member star for EC2, shaded red
(upper panels). The M33 systemic velocity (-179 $\kms$) is shown (dashed line).
For EC1, the best fit halo (green) and disk (red) populations are shown as Gaussian curves, $\sigma_{V,halo}=52 \kms$ at systemic velocity, and $\sigma_{V,disk}=16 \kms$  at -110.3 $\kms$, after removing likely Milky Way stars
(see Trethewey 2011 for details).
For EC2, we show the same Gaussian curves scaled to the stars which are not likely to be Milky Way by velocity and location in the CMD.
In the middle panels, the stars are then shown as a function of radius 
from the centres of the ECs, where half-light radius is only $\sim20$pc  ($\sim5''$) in each case. Photometric 
[Fe/H] as described in the text is shown on the bottom panels, 
revealing the ECs to have likely [Fe/H]$\sim$-1.3, -1.5 respectively, consistent with their RGBs from the CMDs in Huxor et al.\ (2009, 2010). 
\label{ecvels}}
\end{figure}

In Figure~\ref{ecvels}, we show the distribution of velocities of observed stars
in the EC1 and EC2 fields. As expected from Figure~\ref{ecxieta}, the vast
majority of targeted stars lie at large radial distances and are
almost certainly unrelated to the ECs. 
We can in fact only
differentiate the EC stars from the field by plotting the velocities
against their radius from the EC centres (middle panels of Figure~\ref{ecvels}). 
Here, our central targeted
star in each EC is apparent, with velocities of $-152.0\pm4.5
\kms$\ (EC1) and $-200.2\pm5.8 \kms$\ (EC2).  In the case of EC1, a
star lying 15$''$ away ($\sim60$pc -- 3 core radii) is identified as a second
candidate member, by velocity ($-149.3\pm7.2 \kms$) and colour lying
on the identified RGB of Stronkute et al.\ (2008).  
However, adopting the membership probability criteria of Collins et al.\ (2012), this star is unlikely to be a member. Given the possible tidal-disrupted nature of the ECs however, this star may be a former member of the cluster, one that is
no longer bound.
In the EC2 field, no other members are identified.  Finally, the spectroscopically
derived [Fe/H] from the CaT lines is shown as a function of radial
velocity in Figure~\ref{ecvels}, revealing comparable metal-poor [Fe/H]'s
($\sim-1.5$ in all cases) to that published for the two ECs.

\begin{figure}
\begin{center}
\includegraphics[angle=0,width=0.5\textwidth]{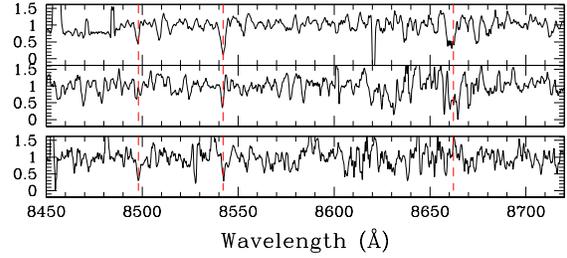}
\vskip-4.6cm
\caption{
The individual spectra of the two  EC1 members (above), and the EC2 member (below), are displayed. 
All three spectra have robust cross correlations with the CaT lines, but as can be seen, the third CaT line is noisy as a result of poor sky subtraction.
}
\label{ecspec}
\end{center}
\end{figure}

In Figure~\ref{ecspec}, we show the individual spectra of the plausible EC1 and
EC2 members, revealing well detected CaT lines and reliable systemic
velocity measurements.
Clearly no dispersion estimates are possible here. It is conceivable
that the bright identified star in the spectrum is actually  a
composite of several cluster stars.  If so, the line width would
provide an indication of the stellar velocity dispersion. However with
$\sim40 \kms$ resolution, and with the spectra clearly dominated by
the brightest targeted star, we cannot hope to resolve any significant
line width (nor do we).
 
The only other extended cluster in the M31 system that has been
studied to date spectroscopically is EC4 (Collins et al.\ 2009),
which clearly resides in the M31 halo.  These authors presented a
spectroscopic survey of candidate RGB stars in the
extended star cluster, \ec4, overlapping two M31 tidal streams.  Six
stars lying on the red giant branch within 2 core-radii of the centre
of \ec4 were found to have an average
$v_{r}=-287.9^{+1.9}_{-2.4}$~kms$^{-1}$ and $\sigma_{v,
  corr}=2.7^{+4.2}_{-2.7}$~kms$^{-1}$, with a resulting mass-to-light
ratio for EC4 of M/L$=6.7^{+15}_{-6.7}\msun/\lsun$, a value that is
consistent with a globular cluster within the $1\sigma$ errors.
Considering several formation and evolution scenarios which could
account for our kinematic and metallicity constraints on EC4, they
conclude that EC4 is most comparable with an extended globular cluster
rather than a dark matter dominated galaxy.  However they find that
both the kinematics and metallicity of EC4 bear a striking resemblance
to M31's tidal stream `Cp' (Ibata et al.\ 2007; Chapman et al.\ 2008), as described in the introduction,
motivating a search for similar tidal structures around EC1 and EC2.


We have searched the regions around EC1 and EC2 for relative
overdensities of stars compared to the disk and halo projected
profiles (Cockcroft et al.\ 2012),
however we find no conclusive evidence for either lying in a clear overdensity. 
EC1 is close enough to the outer disk and stellar tidal tails of M33 that it is difficult to make an assessment.
The 0.5~deg$^2$ field surrounding EC2 is within 1$\sigma$ the density of the halo profile estimate of Cockcroft et al.\ (2012).

In velocity space, we also do not find any obvious evidence for sharp, kinematic stream-like structure (cold velocity spikes) around either EC1 or EC2 (as was found for M31-EC4), except for the second star at 3 core radii in EC1 discussed above. 
Figure~\ref{ecvels} also shows that neither of the ECs is rotating with the disk population, which is  V$_{\rm helio}$$\sim$-110 $\kms$\ at the location of EC1 (uncorrected for disk inclination), with $\sigma_{V,disk}=16 \kms$ after  removing likely Milky Way stars, comparable to that found in McConnachie et al.\ (2006) and Trethewey  (2011),
suggesting that they are true satellites within the M33 halo. 
For EC1, the best fit halo and disk populations are shown as Gaussian curves, $\sigma_{V,halo}=52 \kms$ at systemic velocity, and $\sigma_{V,disk}=16 \kms$  at -110.3 $\kms$, after removing likely Milky Way stars
(see Trethewey 2011 for details).
In EC2  it is not clear that the disk is still kinematically identified from the 
MW foreground (Trethewey 2011), and we show the same Gaussian curve scaled to the stars  not likely to be  foreground (identified by velocity and location in the CMD).

\begin{figure*}
\centering
\includegraphics[width=0.85\textwidth, angle=0]{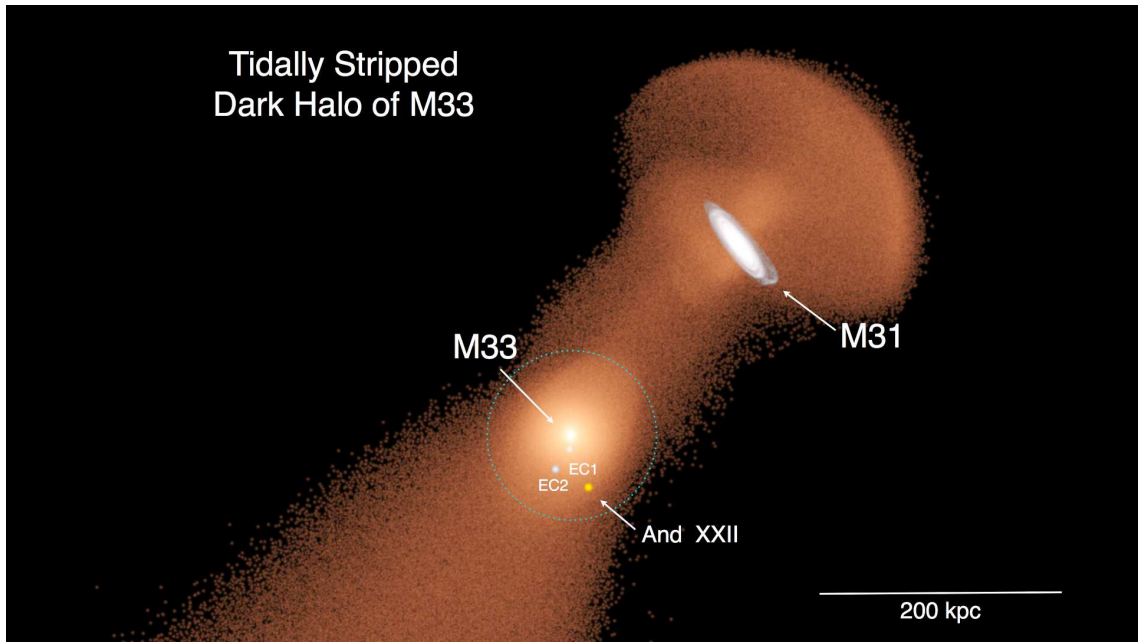}
\caption{
A model of the distribution of dark matter for M33 after an interaction with M31.
The initial dark matter halo of M33 was tidally truncated at a radius
of 70 kpc with a total mass of $1.1 \times 10^{11} {\rm M}_\odot$.  
The disk and bulge of the M31 n-body model is shown for
reference though the dark halo of M31 is intentionally not shown to reveal
the expected distribution of M33 halo mass after the interaction.
We also draw a circle of radius 70 kpc around M33 representing the 
tidal radius in this model.  After the encounter, approximately 40\% of the
initial dark halo mass of M33 is stripped off and either falls back onto
M31 or is flung towards the SW direction.  
The current positions of \andxxii, EC1 and EC2 are shown in their observed
relative positions to M33 in the model.  \andxxii\ lies near the edge of the tidal
radius in projection in this post-interaction model of the halo of M33
while EC1 and EC2  still appear to be in the halo.  An animation of 
the rotation of the model shows that each of \andxxii, EC1 and EC2 are near
the tidal radius of the stripped M33 halo
(http://www.ast.cam.ac.uk/$\sim$schapman/and22.mov).
If this scenario is correct, these objects were probably more tightly bound 
to M33 prior to the interaction and are either just barely bound now 
or are just about to escape from M33 due to tidal stripping.
\label{a22sim}}
\end{figure*}

However, EC1 does seem to reside within a cloud of apparent `halo' stars in M33. A kinematic `halo' component was first discovered in M33 from the spectroscopic survey in McConnachie et al.\ (2006). In the EC1 field, this halo component is well detected and peaks near the M33 systemic velocity of -179 km/s, with a best fit $\sigma_{V,halo}=52 \kms$.
Trethewey (2011) have demonstrated that the `halo stars' in   the EC1 field represent a clear overdensity relative to the radial profile of this halo component (derived from the stars centred about the systemic velocity in 14 other spectroscopic fields 
along the major axis of M33).
However, this relatively strong M33 `halo' component ($\sim$5\% of the disk density) may more likely be a distorted or disrupted disk component as result of the encounter with M31, or even a very high dispersion thick disk component identified as in M31 (Collins et al.\ 2011b). This hypothesis is strengthened by the recently discovered   stellar halo in M33  identified photometrically (Cockcroft et al.\ 2012), which  appears substantially weaker ($\sim1$\% of the disk density) than the spectroscopically identified `halo'  population centred about the systemic velocity.

\section{Discussion}


In this section we address the question of whether \andxxii\ is a satellite
of M33.
As we will see, our distance and radial velocity determinations
play a key role in the discussion.  We also consider constraints on the
mass of M33 from the known satellites in its outer halo.

\subsection{Simulation of the M31-M33 interaction}

As part of the PAndAS project, we have run a series of simulations to
explore the hypothesis that M33 passed within $\leq 60\,{\rm kpc}$
of M31 at some stage during the past few Gyrs.  This hypothesis was
proposed to explain the existence of stellar debris in the outskirts
of the M33 disk (McConnachie et al.\ 2009, Dubinski et al., in
preparation).  

The initial conditions for M31 and M33 are generated using the
GalactICS code (Kuijken \& Dubinski 1995, Widrow, Pym, \& Dubinski
2008) and comprise three collisionless components: an exponential
disk, a Sersic bulge, and an extended dark matter halo.  The models
generated by this code are specified in terms of the phase space
distribution functions (DFs) for the different components.  The DFs in
turn are built out of functions of the integrals of motion and
therefore, through Jeans theorem, provide equilibrium solutions.  In
the current version of the model, it is assumed that the velocity
distribution of the halo particles is isotropic and that their DF
depends only on the energy.

The density profile for the halo is given by \begin{equation} \rho(r)
= \frac{\rho_0}{r/a\left (1 + r/a\right )^\beta}T(r_{\rm out},\delta
r_{\rm out}) \end{equation} where $a$ and $\rho_0$ are, respectively,
the scale length and density parameter for the halo while $\beta$
controls the power-law fall-off of the density at large radii.  The
function $T$ varies smoothly from $1$ to $0$ and has the effect of
truncating the halo at a radius $\sim r_{\rm out}$ over a range in
radii of a few $\delta r_{\rm out}$.  The distribution function for
the halo is determined from the density profile via as Abel transform.

The structural parameters of the different components are adjusted so
that the composite models match the surface brightness profiles and
rotation curves for the two galaxies.  For M31, we use the composite
rotation curve from Widrow:2003 and the surface brightness profile
from Walterbos:1987 while for M33, we use the rotation curve from
Corbelli:2003 and the surface brightness profile from Regan:1994.  We
follow cosmological arguments from Loeb et al.\ (2005) and tune the
M31 model parameters so that the halo mass within 280\,{\rm kpc} is
$M=2.5\times 10^{12}$ $\msun$.  The key structural parameters for the
two galaxies are given in Table 1 of McConnachie:2009.

Candidate orbits for M33 are found by first modelling M33 as a test
particle that moves under the influence of the M31 gravitational field
and dynamical friction.  The latter is incorporated in the calculation
via the Chandrasekhar formula where the Coulomb logarithm is
calibrated using N-body simulations.  We search the space of orbits to
find initial conditions that lead to a strong encounter between M33
and M31 and also to a present-day position and velocity of M33 that
are consistent with observational constraints.  Representative
candidate orbits are then simulated using standard N-body methods.
The simulations are fully self-consistent in the sense that all
components of both M31 and M33 are ``live'' (i.e., the potentials
evolve).  Both M33 and M31 halos were simulated with 2M particles.

For M33, the tidal radius at the initial position of a typical
candidate orbit is $\approx 70$ kpc and so we set parameters of the
function $T$ so that the halo is truncated at that radius.  We also
ran a 50~kpc truncation radius model and the results are similar.  Our
M33 model has a halo mass of $1.1 \times 10^{11}$ solar masses.  The
modeling details are laid out in Dubinski et al.\ (in prep).

In this paper, we focus on a particular example in which M33 passes
within $50\,{\rm kpc}$ of the M31 centre.  The distribution of dark
matter for M33 after an interaction with M31 is shown in
Fig.~\ref{a22sim}.  The disk and bulge of the M31 n-body model is
shown for reference though the dark halo of M31 is intentionally not
shown to reveal the expected distribution of M33 halo mass after the
interaction.  After the encounter approximately, 40\% of the initial
dark halo mass of M33 is stripped off and either falls back onto M31
or is flung towards the SW direction. 

\subsection{Is \andxxii\ a Satellite of M33?} 

The assumption that satellites trace dark matter deserves careful
scrutiny.  In the hierarchical clustering scenario, the dark matter halos
associated with galaxies such as M31 grow by accreting smaller subhalos.
Though most of the subhalos are tidally disrupted, some survive and a   
fraction of these will harbour the satellite galaxies observed today.
Thus, the spatial distribution of satellite galaxies vis-a-vis the dark  
matter depends on the distribution of surviving subhalos and the  
distribution of satellites in comparison with the subhalos.

Numerical simulations of structure formation in a pure dark matter
universe suggest that subhalos are less concentrated than the dark matter
(Ghigna et al.\ 2000, Gao et al.\ 2004, Diemand et al.\ 2004, DeLucia et al.\ 2004), 
presumably because  the subhalos in the inner regions of the parent halo are more susceptible
to tidal disruption.  On the other hand, satellites are more robust than
their pure dark matter counterparts (Nagai et al.\ 2005, Libeskind et al.\ 2010).  These
results may explain why Sales et al.\ 2007 find that the spatial distribution and
kinematics of the satellites and dark matter in their cosmological
simulations are very similar.

In addressing the question of whether \andxxii\ is a satellite of M33, we make the assumption that satellites
trace dark matter.  Therefore, the ratio of the phase space density of
M33 halo particles at a given position and velocity to that of M31
provides an estimate of the probability that a satellite with these
phase space coordinates is a satellite of M33 relative to the
probability that it is a satellite solely of M31.  (Since M33 is very
likely a satellite of M31, any satellite of M33 is also a member of
the M31 system.)  Of course, in calculating this probability, we must
marginalize over unknown quantities such as the proper motion of
\andxxii.

\begin{figure}
\centering
\includegraphics[width=0.65\textwidth, angle=0]{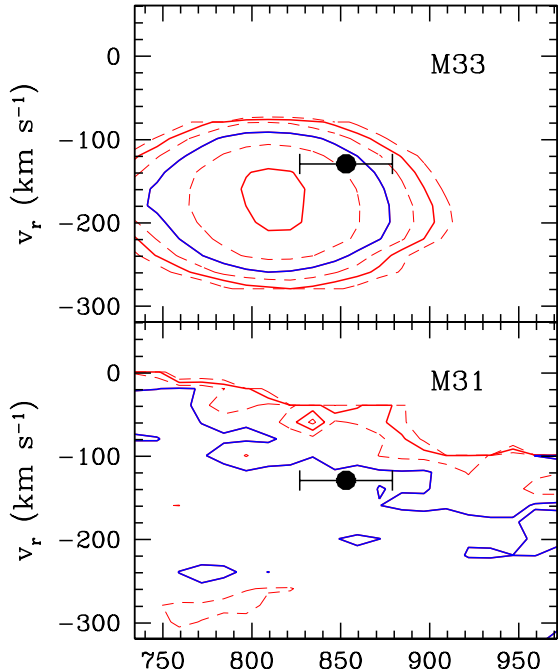}
\vskip-0.5cm
\caption{Phase space density, v$_{\rm r}$ (km s$^{-1}$) versus d$_{\rm r}$ (kpc),
of halo particles in M33 (top panel) and
  M31 (bottom panel).  The upper panel gives the density for particles
  that are initially part of the M33 halo and remain so through the
  interaction with M31.
Solid contours are spaced by 1 dex (dashed are at 0.5 dex).  Levels
are the same in both panels, with the blue line (drawn to go through
And\,XXII in the M33 halo density) providing a reference between the
two.
A comparison of the two at the phase space position of And\,XXII
(black filled circle with error bars on radial velocity and
HB-distance) gives the probability for And\,XXII to be a satellite of
M33 (and thus also a satellite of M31), as compared with the
probability of it being only an M31 satellite, not associated with
M33.  We account for the uncertainty in the M31 and M33 distances
(both 19 kpc -- Conn et al.\ 2012), which spreads out the density
distribution in distance ${\rm d_r}$, especially for M33.  The M31
halo has almost 1\,dex lower density than M33 at this point, although
the uncertainty in distance to And\,XXII has a larger effect on the
relative density in M33 than M31.
\label{a22phase}}
\end{figure}

\begin{figure}
\centering
\includegraphics[width=0.53\textwidth, angle=0]{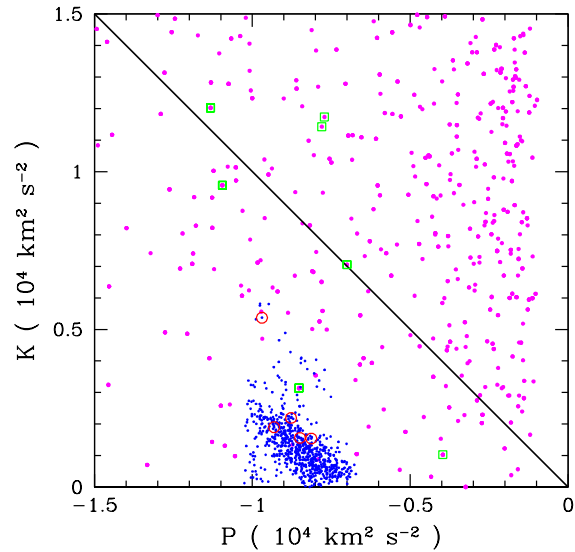}
\vskip-1.1cm
\caption{ In this plot, we take the particles along the And\,XXII line
  of sight (LOS) and calculate their kinetic and potential energy with
  respect to M33.  That is, $K = 0.5 s^2$ where $s$ is the speed of a
  particle in the M33 frame of rest and $P = - {\rm potential\ energy}$
  with respect to M33 particles.
 Magenta (blue) points are M31 (M33) halo particles in a cone along the LOS to
 And\,XXII,  Green squares (Red circles)  are the M31 (M33) particles within 2$\sigma$
 error bars of our \andxxii\ HB distance and velocity values, while the solid
 line divides bound and unbound particles.
%
\label{a22energy}}
\end{figure}

In Fig.~\ref{a22sim}, the current positions of
\andxxii, EC1 and EC2 are shown in their observed relative positions
to M33 in the model.  \andxxii\ can be seen to lie near the edge of
tidal radius in projection in this post-interaction model of the halo
of M33 while EC1 and EC2 still appear to be in the halo.  An animation
of the rotation of the model shows that each of \andxxii, EC1 and EC2
are near the tidal radius of the stripped M33 halo.  If this scenario
is correct, these objects were probably more tightly bound to M33
prior to the interaction and are either just barely bound now or are
just about to escape from M33 due to tidal stripping.  If M33 didn't
have the strong interaction implied by this simulation, then its halo
would be more extended making \andxxii\ even more likely to be a
satellite.

Figure~\ref{a22phase} shows a reduced phase space density of particles
close to the \andxxii\ line-of-sight that are initially part of either
the M33 halo or M31 halo.  The reduced phase space density corresponds
to the number of stars per unit line-of-sight distance and unit radial
velocity.  We first select stars from either the M31 or M33 halo that
lie within 0.01 rad of the line-of-sight to \andxxii.  These stars are
then binned in terms of $d_r$ and $v_r$.  The number of stars in each
bin divided by the phase space volume of the bin gives the desired
density.  We incorporate uncertainties in the distances to M31 and M33
using a simple Monte Carlo scheme whereby the above procedure is
repeated for different choices of these distances, as prescribed by
the quoted uncertainties.

We see that at the measured coordinates of \andxxii, the M31 halo
density is almost 1\,dex lower  than the M33 halo density.  The
implication is that \andxxii\ is very likely associated with M33. While
the uncertainty in distance to \andxxii\ has a larger effect on the
inferred density in M33 than in M31, the change to the probability is not
large.  Within the $1\sigma$ distance uncertainties the M33/M31
density contrast never drops significantly below 1~dex.
Nevertheless, improvements in the distance determination to \andxxii\
will have the largest effect on resolving the question of whether it
is a satellite of M33.
A similar analysis of the extended clusters, EC1 and EC2, show the phase space densities of the M31 and M33 halos in the vicinities of these objects are comparable to \andxxii\ within factors of a few (although the large radial distance of EC1 makes it the least likely of the three to be a satellite), thus
all three objects are good candidates to be associated with M33.

\begin{figure}
\centering
\includegraphics[width=0.60\textwidth, angle=0]{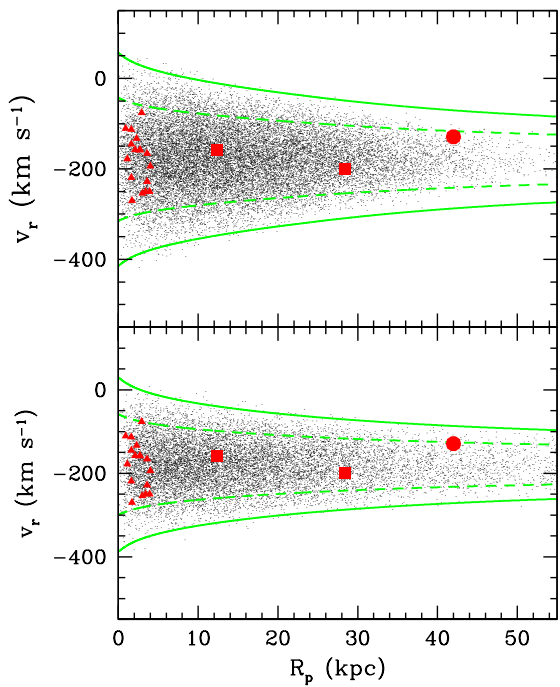}
\vskip-0.5cm
\caption{Distribution of M33 objects in terms of the projected
radius and radial velocity. \andxxii\ is shown as the large blue
circle while EC1 and EC2 are shown as the blue squares.  Blue
triangles show the positions of all {\it halo} globular clusters (GCs)
from Chandar et al.\ (2002) (i.e., GCs with $|V_{\rm disk}-V_{GC}| > 70
\kms$).  Black dots are from the simulations where the top panel is
for initial conditions and the bottom panel is for the present-day
distribution.  The solid red curves give an estimate of the escape
speed from M33 shifted to the Galactic frame, $v_{\rm esc}-179\,\kms$
where $v_{\rm esc} \equiv \sqrt{2\phi}$ and $\phi$ is the spherical
approximation to the gravitational potential.  The dashed red curves
use $v_{\rm esc}/\sqrt{3}$ to roughly account for the fact that the
radial velocity is one of three velocity components.  Note that the
escape speed decreases with time due to tidal stripping of M33 during
its encounter with M31.
\label{m33mass}}
\end{figure}

In Figure~\ref{a22energy}, we plot the particles close to the \andxxii\ line-of-sight
(LOS) as a function of their kinetic energy K and binding energy P, where 
these quantities are calculated in the rest frame of M33.
 The diagonal line in the figure divides bound and unbound particles
 and reveals firstly that in the direction of And\,XXII, all of the
 particles initially in the M33 halo remain bound to M33, and secondly
 that some of the M31 particles become bound to M33. 
  Further analysis
 is required to determine the extent to which M33 can sweep up M31
 halo particles.  

In McConnachie et al.\ (2009) we provided an animation of an M31-M33
encounter that included an extended stellar halo for M33.  This
animation showed that a significant number of halo particles were
stripped from M33.  The point here is that the particle distribution
in the cone along the \andxxii\ line-of-sight is dominated by material
that is still bound to M33 and that only a few particles are caught in
the act of being tidally stripped.  Most of the stripped particles seen
in the simulation are at projected distances from M33 much larger
than the projected radius of \andxxii.


\subsection{Constraining the mass of M33 with its satellites}


In Figure~\ref{m33mass} we compare the projected radius and radial velocity for
\andxxii, EC1 and EC2 with the escape velocity, $v_{\rm esc} \equiv
\pm \sqrt{2|\phi|}$ as derived from our M33 model, where $\phi$ is the spherical
approximation to the gravitational potential.  Also shown are
curves $v_{\rm esc}/\sqrt(3)$ where the factor of $1/\sqrt{3}$ is
included to account for the fact that the radial velocity is one of
three components.  Figure~\ref{m33mass}  contrasts the initial conditions with the present-day
distribution, where the
escape speed decreases with time due to tidal stripping of M33 during
its encounter with M31.
The projected distance represents a lower
bound on the true physical distance of an object to the M33 centre.
In fact, the animation in Figure~\ref{a22sim} reveals that each of \andxxii, EC1 and EC2 is near
the tidal radius of the stripped M33 halo at their true distances, with 
EC1 being the most distant satellite of the three at 71.2~kpc.
All three objects have radial velocities relative to M33 that are
below the apparent escape speed, emphasizing further that they are very likely bound satellites, 
although \andxxii\ comes close to the $v_{\rm esc}/\sqrt{3}$. 
These results suggest that the fiducial
dynamical mass estimates for M33 ($\sim10^{11}$ M$_\odot$) are
consistent with the M33 outer satellite system.  By contrast, M31
appears to have several satellites that have radial velocities at or
above the nominal escape velocity (Chapman et al.\ 2007).

Cockcroft et al.\ (2011) extended previous studies of Globular Clusters (GCs) in M33
(Huxor et al.\ 2009) out to a projected radius of 50~kpc and covers
over 40 square degrees. They found only one new unambiguous star
cluster (at a projected radius of 22~kpc) in addition to the five
previously known in the M33 outer halo (ranging from 9~kpc to 28~kpc).
In total then, other than these three objects, there are only 3 other
satellites (all GCs) in the 10-50 kpc halo of M33.  To fill in the
inner region of the plot, we also show the `halo' GCs from Chandar et
al.\ (2002)
defined to have  $|V_{disk}-V_{GC}| > 70$ $\kms$.

Figure~\ref{m33mass}  also provides a visual depiction of the evolution of M33
halo particles and the relation between \andxxii\ and the halo
particles. The effect of the encounter is to populate regions with low
binding energy, as expected. The simulation also shows that the curves
illustrating the escape velocity of the M33 potential are a reasonable
approximation to the projected 3D properties of the bound satellite
GCs and dSph.  In the panel depicting the system after the encounter,
there is a much higher density of halo particles bound to M33 at the
position of \andxxii.  While \andxxii\ may well be a satellite of M33,
its phase space position was likely affected by M31 during the M33-M31
encounter.






\section{Conclusions}

We have explored the association between M33 and three possible satellites, \andxxii, EC1, and EC2, finding all three were likely to have originally been bound to M33, and likely still are, depending on the nature of the interaction of M31--M33.  
From a spectroscopic survey of these candidate satellites,
we have defined probable members, for which we derive radial velocities down to $i$$\sim$23,
and with average errors of 5$\kms$.
The \andxxii\ heliocentric velocity  is v$_r$ =-130.0$\pm{1.7} \kms$, 
+177~km/s relative to M31, but only +50~km/s relative to M33 (in agreement with Tollerud et al.\ 2012, v$_r$ =-126.8$\pm3.1\kms$).
The
dispersion is unresolved with 10  member stars at
$\sigma_v < 6.0 \kms$, 99.5\% confidence.
Using the photometry and spectroscopy of the confirmed member stars, we find a metallicity with median [Fe/H]$=-1.6\pm0.1$.  
Our identified member stars in \andxxiii\ along with an assessment of the horizontal branch peak suggest a radial 
distance of 853$\pm$26~kpc. 



For the two extended clusters in M33, we also find systemic velocities of $-152.0\pm4.5 \kms$\ (EC1)   and $-200.2\pm5.8 \kms$\ (EC2).
EC1 is further found to lie in a local kinematic overdensity of apparent halo stars in M33, suggesting it may be a signpost for a recently disrupted dSph satellite of M33.

The combined velocity and distance information is used to assess whether \andxxii, EC1, and EC2 could have been, or are still  bound satellites of M33 using our simulations of the M31--M33 interaction. 
We conclude that all three are highly  likely  to have originally been satellites of M33.
All three satellites have radial velocities relative to M33  below the
apparent  escape speed, and modelling of a possible M33-M31 interaction indicates that all three satellites lie near the model virial radius, $\sim70$~kpc, 
suggesting they could be close to the point of being bound to M33 if  tangential velocities were comparable to radial velocities. If M33 didn't have the strong interaction implied by our simulation, then its halo would be more 
extended making the \andxxii, EC1, and EC2  even more likely to be satellites.
The fiducial dynamical mass estimates for M33 ($\sim10^{11}$ M$_\odot$) are consistent with the M33 outer satellite system.

\section*{ACKNOWLEDGMENTS}
SCC thanks the STFC for financial support. RAI  gratefully acknowledges support from the Agence Nationale de la Recherche though the grant POMMME (ANR 09-BLAN- 0228). GFL  thanks the Australian Research Council for support through his Future Fellowship (FT100100268) and Discovery Project (DP110100678).
JP acknowledges support from the Ram\'on y Cajal Program as well as by the
Spanish grant AYA2010-17631 awarded by the Ministerio of Econom\'ia y
Competitividad.
The data presented herein were obtained at the W.M.
Keck Observatory, which is operated as a scientific partnership among
the California Institute of Technology, the University of California and
the National Aeronautics and Space Administration. The Observatory was
made possible by the generous financial support of the W.M. Keck
Foundation. 
Based on observations obtained with MegaPrime/MegaCam, a joint project of CFHT and CEA/DAPNIA, at the Canada-France-Hawaii Telescope (CFHT) which is operated by the National Research Council (NRC) of Canada, the Institute National des Sciences de l'Univers of the Centre National de la Recherche Scientifique of France, and the University of Hawaii.

\begin{table*}
\begin{center}
\caption{Properties of candidate member stars in \andxxii\ centered at $\alpha$ = 01h 27m 39.6s, $\delta$ = 28$^\circ$ 05$'$ 29$"$.}
\begin{tabular}{@{\extracolsep{-1.5pt}}lllcccccc}
\noalign{\medskip}
\hline
$ID^a$ & $\alpha$ (J2000) & $\delta$ (J2000) & $i-mag$ & $g-i$ & D$_{rad}^b$ & [F e/H]$_{phot}^c$  & v$_r$ ($\kms$) & probability$^d$ \\
\hline
57 & 01 27 41.10 & 28 05 34.4 & 21.65 & 1.23 &  1.11 &  -1.58 & -130.7 $\pm$ 3.7 & 0.958\\
76 & 01 27 37.34 & 28 05 54.4 & 23.24 & 1.12 &  0.76 &  -1.78 & -119.0 $\pm$ 16.9 & 0.774\\
42 & 01 27 40.26 & 28 05 20.2 & 22.88 & 1.05 &  0.10 &  -1.87 & -131.9 $\pm$ 6.9 & 0.737\\
24 & 01 27 43.63 & 28 04 52.0 & 23.34 & 1.07 &  0.97 &  -1.91 & -129.4 $\pm$ 8.3 & 0.698\\
60 & 01 27 35.48 & 28 05 37.2 & 22.15 & 1.24 &  1.02 &  -1.02 & -125.6 $\pm$ 5.4 & 0.666\\
29 & 01 27 47.09 & 28 04 58.5 & 23.02 & 1.04 &  1.63 &  -1.40 & -130.8 $\pm$ 9.2 & 0.366\\
77 & 01 27 37.77 & 28 05 54.4 & 21.66 & 1.39 &  0.69 &  -1.13 & -128.6 $\pm$ 3.7 & 0.282\\
95 & 01 27 34.17 & 28 06 30.9 & 22.29 & 1.26 &  1.69 &  -1.58 & -129.5 $\pm$ 5.4 & 0.251\\
53 & 01 27 35.21 &  28 05 29.9 & 20.87 & 0.65 &  1.15 &     ---  & -113.2 $\pm$ 1.9  & 0.010\\
86 & 01 27 55.63 & 28 06 18.0 & 23.04 & 1.23 &  3.56 &  -1.68 & -133.9 $\pm$ 9.3 & 0.003\\
10 & 01 27 23.05 & 28 04 12.6 & 22.64 & 0.76 &  3.93 &  --- & -130.4 $\pm$ 7.8 & 0.001\\
\hline
6$^e$ & 01 27 42.66 & 28 04 1.6 & 20.42 & 0.63 &  1.55 &  --- & -155.2 $\pm$ 7.8 & $<$10$^{-4}$\\
131$^f$ & 01 28 20.16 & 28 08 43.0 & 20.78 & 1.29 &  9.91 &  --- & -111.5 $\pm$ 7.8 & $<$10$^{-4}$\\
\hline
\hline
\end{tabular}
\end{center} 
\noindent$^a$ Stars are ordered by their membership probability (see below)\\
$^b$ Radial distance in arcmin from the centre of \andxxii.\\
$^c$ Photometric [Fe/H] defined by fitting Dotter isochrones to the $I,g$-mags, as described in the text. ID\#10 lies off the range of isochrones, but is at sufficiently large distance from \andxxii\ that its probability of membership is miniscule, and it has no effect on the dispersion or systemic velocity estimates.\\
$^d$ Candidate member stars are assigned a probability of membership to \andxxii,  defined by a combination of velocity, distance from the RGB locus, and radial distance from the centre of \andxxii\ (see Collins et al.\ 2012 for details).\\
$^e$ This star is not considered as a member star with it's low probability based on both CMD position and velocity offset from systemic. It is likely an M33 halo star.\\
$^f$  This star is not considered as a member star with it's low probability based mainly on its 10$'$ distance from the centre of \andxxii.
\end{table*}

\begin{table*}
\begin{center}
\caption{Properties of the candidate member stars in M33's EC1 and EC2,  centered at (EC1) $\alpha$ = 01h 32m 58.5s, $\delta$ = 29$^\circ$ 52$'$ 03$"$, (EC2) $\alpha$ = 01h 35m 41.8s, $\delta$ = 28$^\circ$ 49$'$ 16$"$. 
}
\begin{tabular}{@{\extracolsep{-1.5pt}}lllccccc}
\noalign{\medskip}
\hline
$ID$ & $\alpha$ (J2000) & $\delta$ (J2000) & $i-mag$ & $g-i$ & D$_{rad}$ & [F e/H]$_{spec}$  & v$_r$ ($\kms$) \\
\hline
EC1: \\
\hline
20 & 01 32 57.91 &  29 52 16.9 & 23.02 & 1.04 &  0.25 &  -1.48 & -149.3$\pm$7.2\\
127 & 01 32 58.50 & 29 52  02.9 & 21.65 & 1.23 &  0.0 &  -1.59 & -152.0$\pm$4.5\\
\hline
EC2:\\ 
\hline
70 & 01:35:41.78  & 28:49:15.5 & 21.04 & 1.15 & 0.0 & -1.54 & -200.2$\pm$5.8\\
\hline
\hline
\end{tabular}
\end{center} 
\end{table*}

\newcommand{\mnras}{MNRAS}
\newcommand{\pasa}{PASA}
\newcommand{\nat}{Nature}
\newcommand{\araa}{ARAA}
\newcommand{\aj}{AJ}
\newcommand{\apj}{ApJ}
\newcommand{\apjl}{ApJ}
\newcommand{\apjs}{ApJSupp}
\newcommand{\aap}{A\&A}
\newcommand{\aaps}{A\&ASupp}
\newcommand{\pasp}{PASP}

\end{document}